\documentclass[10pt,twocolumn,letterpaper]{IEEEtran}
\usepackage{flushend} 
\usepackage{graphicx}
\usepackage{epsfig}
\usepackage{latexsym}
\usepackage{amsfonts}
\usepackage{here}
\usepackage{rawfonts}
\usepackage[latin1]{inputenc}
\usepackage[T1]{fontenc}
\usepackage{calc}
\usepackage{url}
\usepackage{enumerate}
\usepackage{color}
\usepackage{amssymb}
\usepackage{bm}
\usepackage{upref}
\usepackage{times}
\usepackage{amsmath}
\usepackage{cite}

\newcommand{\ms}{\;\;}

\newcommand{\mycaption}[1]{\vspace*{-1.6em}\caption{#1}\vspace*{-1.0em}}

\newtheorem{lemma}{{\bf Lemma}}

\newcommand{\qed}{\nobreak \ifvmode \relax \else
  \ifdim\lastskip<1.5em \hskip-\lastskip
  \hskip1.5em plus0em minus0.5em \fi \nobreak
  \vrule height0.75em width0.5em depth0.25em\fi}

\newlength{\aligntop}
\newlength{\alignbot}
\makeatletter
\renewenvironment{align}{%
  \vspace{\aligntop}
  \start@align\@ne\st@rredfalse\m@ne
}{%
  \math@cr \black@\totwidth@
  \egroup
  \ifingather@
    \restorealignstate@
    \egroup
    \nonumber
    \ifnum0=`{\fi\iffalse}\fi
  \else
    $$%
  \fi
  \ignorespacesafterend%
  \vspace{\alignbot}\par\noindent
}
\makeatother

\IEEEoverridecommandlockouts

\author{Manav R. Bhatnagar,~\IEEEmembership{Senior Member,~IEEE} 
\vspace*{-2.0em}%
  \thanks{\textbf{Corresponding author:} Manav R. Bhatnagar is with
    Department of Electrical Engineering, Indian Institute of Technology Delhi, Hauz Khas, IN-110016 New Delhi, India,
    email: \protect\url{manav@ee.iitd.ac.in}.}%
}\date{}

\title{On the Capacity of Decode-and-Forward Relaying over Rician Fading Channels\vspace*{-0.15em}}

\begin{document}

\maketitle
\vspace*{-1.0em}
\begin{abstract}
In this letter, 
we derive the probability density function (PDF)
and cumulative distribution function (CDF) of  
the minimum of two 
non-central Chi-square random variables with
two degrees of freedom in terms of power series. 
With the help of the derived PDF and CDF, we obtain
the exact ergodic capacity of the following adaptive protocols in a
{decode-and-forward} (DF) cooperative system over dissimilar 
{Rician} fading channels: 
(i) constant power with optimal rate
adaptation; (ii) optimal simultaneous
power and rate adaptation;  (iii) channel inversion with fixed rate. 
By using the analytical expressions of the capacity, 
it is observed that the optimal power and
rate adaptation provides better capacity than the optimal rate adaptation with constant
power from low to moderate signal-to-noise ratio values over dissimilar Rician fading channels. 
Despite low complexity, the channel inversion based adaptive transmission is
shown to suffer from significant loss in capacity as compared to the
other adaptive transmission based techniques over DF Rician channels.
\end{abstract}
\section{Introduction}
Information theoretic study of a cooperative network is a useful tool for implementation of this technology in
current and future wireless networks. 
There exists many useful works over the capacity analysis of the cooperative networks~\cite{avest07,beaul06,surav06,shres08}. 
The outage performance of the cooperative system with Rayleigh fading in the low signal-to-noise ratio (SNR) regime is studied in~\cite{avest07}. Closed-form expressions of the outage probability over Rayleigh and Nakagami-$m$ fading channels are derived in~\cite{beaul06} and~\cite{surav06}, respectively, with decode-and-forward (DF) relays. The outage capacity performance of the two hop based amplify-and-forward (AF) protocol over dissimilar Rayleigh fading channels is studied in~\cite{shres08}. 

All the aforementioned works only consider fixed rate and fixed power transmission based schemes. 
However, if the source/relay terminal possesses the channel state information (CSI), then it can adapt the transmit power level, symbol/bit rate, constellation size,
coding rate/scheme or any combination of these parameters
in response to the changing channel conditions~\cite{aloui99,nechi08,annam10,farha10}. In~\cite{nechi08}, the ergodic capacity of a two hop AF cooperative system with multiple relays and adaptive source transmission is explored; whereas, capacity of AF multi-hop relaying systems under adaptive transmission is studied in~\cite{farha10}. 
An upper bound of the outage probability of an AF based cooperative system with a single relay in Rician fading channels is derived in~\cite{limpa09}. The capacity of adaptive transmission techniques in a two-hop AF cooperative system with Rician fading is studied in~\cite{modi11}; upper bounds of the capacity are derived in the form of infinite integrals by using moment generating function approach.  However, all these works consider AF protocol; whereas, DF protocol is useful as it requires digital  processing at relay contrary to the AF protocol. 
The capacity of adaptive modulation in a single relay based DF cooperative system over Rayleigh fading channels is studied in~\cite{hasna05}.  
In this paper, by using a well known series representation of the modified Bessel function of the first kind, we use a probability density function (PDF) and cumulative distribution function (CDF) based approach which leads to power series 
expressions of the \emph{exact} ergodic capacity of the adaptive transmission based DF system over dissimilar Rician fading channels. The proposed power series expressions converge for finite number of summation terms and are shown to be useful in giving significant insight of the ergodic capacity of the adaptive transmission techniques in the DF cooperative system over dissimilar Rician fading channels. 

Our main contributions are as follows. i) We derive the PDF and CDF of minimum of two non-central Chi-square random variables (RVs) with two degrees of freedom in terms of converging power series. ii) The \emph{exact} average capacity of a two-hop DF cooperative system with adaptive modulation over dissimilar Rician channels is derived by using these PDF and CDF expressions.  iii) Different adaptive transmission techniques are compared on the basis of these expressions over DF Rician channels. 
\section{Characterization of Minimum of Two Non-Central Chi-Square Random Variables with Two Degrees of Freedom} 
\begin{figure}[t!]\vspace*{-1.0em}
  \begin{center}\hspace*{-1.3em}
    \psfig{figure=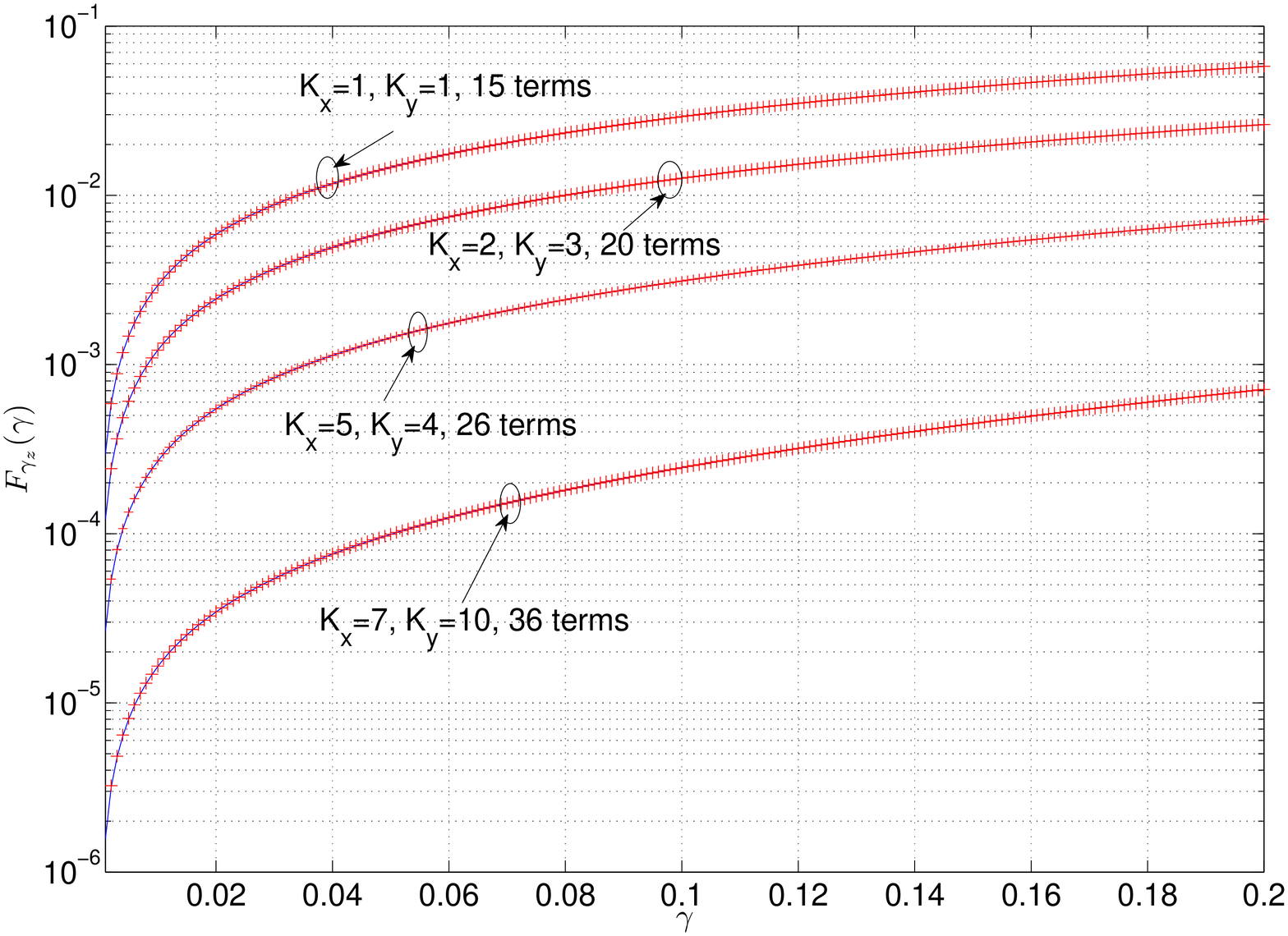,height=2.35in,width=3.95in}
    \vspace*{-1.25em}
    \mycaption{Analytical $+$ and simulated $-\!\!\!-$ CDF of $\gamma_z$ for different values of $K_x$, $K_y$, different number of summation terms, and $\bar{\gamma}_x=\bar{\gamma}_y=5$.}
    \label{fig:cdf}
    \vspace*{-1.2em}
  \end{center}
\end{figure}

Let $G_i\sim\mathcal{N}\left(m_i,\sigma^2/2\right)$, $i=1,2$ be the two Normal distributed RVs with $m_i$ means and $\sigma^2/2$ variance, then $U=G^2_1+G^2_2\sim \mathcal{X}_2\left(K,\bar{\gamma}\right)$, where $K=s^2/\sigma^2$, $\bar{\gamma}=s^2+\sigma^2$, and $s^2=m^2_1+m^2_2$, will be non-central Chi-square distributed with two degrees of freedom as~\cite[Eq.~(2.16)]{simon05} 
\begin{align}
\label{eq:fx}
f_U(u)&=Ae^{-au}I_0\left(2\sqrt{Kau}\right),
\end{align}
where $f_U(u)$ denotes the PDF of $U$, $a=(1+K)/\bar{\gamma}$,  
$A=ae^{-K}$, and $I_0(\cdot)$ denotes the modified Bessel function of the first kind of order zero~\cite{grand07}.  
Let us now define the following RV:
\begin{align}
\label{eq:min}
\gamma_z\triangleq\min\left\{\gamma_x,\gamma_y\right\},
\end{align}
where $\gamma_x\sim\mathcal{X}_2\left(K_x,\bar{\gamma}_x\right)$ and $\gamma_y\sim\mathcal{X}_2\left(K_y,\bar{\gamma}_y\right)$ be independent and non-identically distributed (i.n.i.d.) non-central Chi-square RVs with two degrees of freedom. 
The CDF and PDF of $\gamma_z$ are obtained in the following lemmas which are proved in Appendix~\ref{app:new}. 
\begin{lemma}
\label{lem:new1}
The CDF of $\gamma_z$ is given as
\begin{align}
\label{eq:cdf_min_final}
F_{\gamma_z}(\gamma)=1-A_xA_y\sum^\infty_{k=0}\sum^k_{n=0}&\tilde{B}_x(n)\tilde{B}_y(k-n)\Gamma\left(n+1,a_x\gamma\right)\nonumber\\
&\times\Gamma\left(k-n+1,a_y\gamma\right),
\end{align}
where $a_x=(1+K_x)/\bar{\gamma}_x$, $a_y=(1+K_y)/\bar{\gamma}_y$, $A_x=a_xe^{-K_x}$, $A_y=a_ye^{-K_y}$, $\tilde{B}_x(k)={B}_x(k)/a^{k+1}_x$, $\tilde{B}_y(k)={B}_y(k)/a^{k+1}_y$, $
B_x(k)=\frac{K^k_x(1+K_x)^k}{\bar{\gamma}^k_x\left(k!^2\right)},$ $
B_y(k)=\frac{K^k_y(1+K_y)^k}{\bar{\gamma}^k_y\left(k!^2\right)}$, and $\Gamma(\cdot)$ denotes the Gamma
function. 
\end{lemma}
\begin{lemma}
\label{lem:new}
The PDF of $\gamma_z$ is given as
\begin{align}
\label{eq:pdf_min1}
f_{\gamma_z}(z)=&A_xA_y\sum^\infty_{k=0}\sum^k_{n=0}\tilde{B}_x(n)\tilde{B}_y(k-n)\Big(a^{n+1}_x(k-n)!\nonumber\\
&\times \sum^{k-n}_{m_1=0}\frac{a^{m_1}_{y}}{m_1!}\gamma^{n+m_1}e^{-(a_x+a_y)\gamma}+a^{k-n+1}_y n!\nonumber\\
&\times \sum^n_{m_2=0} \frac{a^{m_2}_x}{m_2!}\gamma^{k-n+m_2}e^{-(a_x+a_y)\gamma}\Big).
\end{align}
\end{lemma}\vspace*{0.25em}
It can be easily shown analytically that \eqref{eq:cdf_min_final} and \eqref{eq:pdf_min1} contain converging power series.
This fact is corroborated by 
Fig.~\ref{fig:cdf} where 
the CDF of $\gamma_z$ 
given in \eqref{eq:cdf_min_final} 
converges nicely to the simulated CDF for finite number of summation terms. 
\section{System Model and Capacity Analysis under Adaptive Transmission}
Let us consider a DF cooperative system with a single \emph{half-duplex} relay and no direct link in between the source and the destination.  
The channels of the source-to-relay and relay-to-destination links are assumed to be i.n.i.d. Rician distributed with $K_x$ and $K_y$ line of sight (LOS) components, and average SNRs $\bar{\gamma}_x$ and $\bar{\gamma}_y$, respectively. Therefore, instantaneous SNRs of the source-to-relay and relay-to-destination links, i.e., $\gamma_x\sim\mathcal{X}_2\left(K_x,\bar{\gamma}_x\right)$ and $\gamma_y\sim\mathcal{X}_2\left(K_y,\bar{\gamma}_y\right)$, respectively, are non-central Chi-square RVs with two degrees of freedom.  

The communication from the source to the destination takes place in two orthogonal channels via the relay that \emph{always} decodes and forwards the data of the source to the destination. 
Hence, both hops act independently
in the sense that decoding/encoding takes place
at the intermediate relay. Consequently, the DF based cooperative transmission is equivalent to a series network, which means that the capacity of the system is dominated by the worst hop. 
Since the capacity is a
monotonous function of SNR, the minimum of the capacities of the source-to-relay and
relay-to-destination channels equals the capacity of the weakest of the source-to-relay
and relay-to-destination channels. Therefore, the equivalent SNR (not end-to-end SNR) of the two-hop DF system from a \emph{capacity point of view} is given in \eqref{eq:min}~\cite{hasna05}. 
In the following subsections, we will derive the capacity of
different adaptive schemes in DF cooperative system with Rician fading channels.
\subsection{Optimal Rate Adaptation with Constant Transmit Power}
 For optimal rate adaptation to the fading level and constant transmit power, the channel capacity of the considered DF relaying system in bits per second is given by~\cite{aloui99}
\begin{align}
\label{eq:optimalratecap}
C=0.5B\int^\infty_{0}\text{log}_2(1+\gamma)f_{\gamma_z}(\gamma)d\gamma,
\end{align}
where $B$ (in hertz) is the bandwidth of the channel. 
Let us now state the following lemma whose proof is given in Appendix~\ref{app:1}. 
\begin{lemma}
\label{lem:1}
For $\alpha\in \mathbb{Z}^*$ and $\beta>0$, we have
\begin{align}
\label{eq:lem1res}
&\int^\infty_0 w^\alpha e^{-\beta w} \text{ln}(1+w) dw\nonumber\\
&=\sum^\alpha_{q=0}{}^\alpha C_q (-1)^{\alpha-q}\frac{e^\beta}{\beta^q}G^{3,0}_{2,3}\left(\beta\Big|\begin{array}{c}\hspace*{-1em}0,0\\-1,-1,q \end{array}\right),
\end{align}
where $G^{m,n}_{p,q}\left(v\Big|\begin{array}{c}a_1,..,a_p\\b_1,..,b_q \end{array}\right)$ is the Meijer's function, $\text{ln}\left(\cdot\right)$ denotes the natural logarithm, and $^\mu C_\nu$ denotes the binomial coefficient. 
\end{lemma}
By substituting $f_{\gamma_z}(\gamma)$ from \eqref{eq:pdf_min1} in \eqref{eq:optimalratecap} and then using Lemma~\ref{lem:1}, we get the capacity of the DF relating with rate adaptation as follows:
\begin{align}
\label{eq:optimalratecap1}
&C=0.5BA_xA_y\text{log}_2e\!\!\sum^\infty_{k=0}\sum^k_{n=0}\!\!\tilde{B}_x(n)\tilde{B}_y(k-n)\Bigg(a^{n+1}_x(k-n)!\nonumber\\
&\times\sum^{k-n}_{m_1=0}\frac{a^{m_1}_y}{m_1!}\sum^{n+m_1}_{q_1=0}{}^{n+m_1}C_{q_1}(-1)^{n+m_1-q_1}\frac{e^{a_x+a_y}}{(a_x+a_y)^{q_1}}\nonumber\\
&\times G^{3,0}_{2,3}\left(a_x+a_y\Big|\begin{array}{c}\hspace*{-1em}0,0\\-1,-1,q_1 \end{array}\right)+ a^{k-n+1}_y n!\sum^{n}_{m_2=0}\frac{a^{m_2}_x}{m_2!}\nonumber\\
&\times \sum^{k-n+m_2}_{q_2=0}{}^{k-n+m_2}C_{q_2}(-1)^{k-n+m_2-q_2}\frac{e^{a_x+a_y}}{(a_x+a_y)^{q_2}}\nonumber\\
&\times G^{3,0}_{2,3}\left(a_x+a_y\Big|\begin{array}{c}\hspace*{-1em}0,0\\-1,-1,q_2 \end{array}\right)\Bigg).
\end{align}
\subsection{Optimal Simultaneous Power and Rate Adaptation}
Given an average transmit power constraint, the channel capacity of a fading channel with optimal power and rate adaptation is given as~\cite{aloui99} 
\begin{align}
\label{eq:optimalpowerandoptimalratecap}
C=0.5B\int^\infty_{\gamma_0}\text{log}_2(\frac{\gamma}{\gamma_0})f_{\gamma_z}(\gamma)d\gamma,
\end{align}
where $\gamma_0$ is the optimal cutoff SNR level below which the data transmission is suspended. This optimal cutoff SNR must satisfy the following relation:
\begin{align}
\label{eq:optcutoff}
\int^\infty_{\gamma_0}\left(\frac{1}{\gamma_0}-\frac{1}{\gamma}\right)f_{\gamma_z}(\gamma)d\gamma=1.
\end{align}
Let us now state the following lemma whose proof is given in Appendix~\ref{app:2}. 
\begin{lemma}
\label{lem:2}
For $\alpha\geq 0$, $\beta>0$, and $\eta>0$, we have
\begin{align}
\label{eq:lem2res}
\int^\infty_\eta w^\alpha e^{-\beta w} \text{ln}(\frac{w}{\eta}) dw=\frac{\eta}{\beta^{\alpha}}G^{3,0}_{2,3}\left(\beta\eta\Big|\begin{array}{c}\hspace*{-1em}0,0\\-1,-1,\alpha \end{array}\right).
\end{align}
\end{lemma}
By substituting $f_{\gamma_z}(\gamma)$ from \eqref{eq:pdf_min1} in \eqref{eq:optimalpowerandoptimalratecap} and then using Lemma~\ref{lem:2}, we get the capacity of DF system with transmit power and rate adaptation over dissimilar Rician fading hops as
\begin{align*}
&C=0.5BA_xA_y\text{log}_2e\!\!\sum^\infty_{k=0}\sum^k_{n=0}\!\!\tilde{B}_x(n)\tilde{B}_y(k-n)\Bigg(a^{n+1}_x(k-n)!\nonumber\\
&\times\sum^{k-n}_{m_1=0}\frac{a^{m_1}_y\gamma_0}{m_1!(a_x+a_y)^{n+m_1}}\end{align*}
\begin{align}
\label{eq:optimalpowerandratecap1}
&\times G^{3,0}_{2,3}\left(\left(a_x+a_y\right)\gamma_0\Big|\begin{array}{c}\hspace*{-1em}0,0\\-1,-1,n+m_1 \end{array}\right)+ a^{k-n+1}_y n!\nonumber\\
&\times \sum^{n}_{m_2=0}\frac{a^{m_2}_x\gamma_0}{m_2! (a_x+a_y)^{k-n+m_2}}\nonumber\\
&\times G^{3,0}_{2,3}\left(\left(a_x+a_y\right)\gamma_0\Big|\begin{array}{c}\hspace*{-1em}0,0\\-1,-1,k-n+m_2 \end{array}\right)\Bigg).
\end{align}\vspace*{-0.7em}

\hspace*{-1em}The optimal cutoff SNR $\gamma_0$ is found by solving for $\gamma_0$ in
\eqref{eq:optcutoff}, which can be rewritten as\vspace*{0.3em}
\begin{align}
\label{eq:optcutoff1}
\frac{1}{\gamma_0}\int^\infty_{\gamma_0}f_{\gamma_z}(\gamma) d\gamma-\int^\infty_{\gamma_0}\frac{1}{\gamma}f_{\gamma_z}(\gamma)d\gamma=1.
\end{align}\vspace*{-0.7em}

\hspace*{-1em}Since $\int^\infty_{\gamma_0}f_{\gamma_z}(\gamma)d\gamma=1-F_{\gamma_z}(\gamma_0)$,  from \eqref{eq:cdf_min_final}, we get
 \begin{figure}[t!]\vspace*{-1.0em}
  \begin{center}\hspace*{-2.3em}
    \psfig{figure=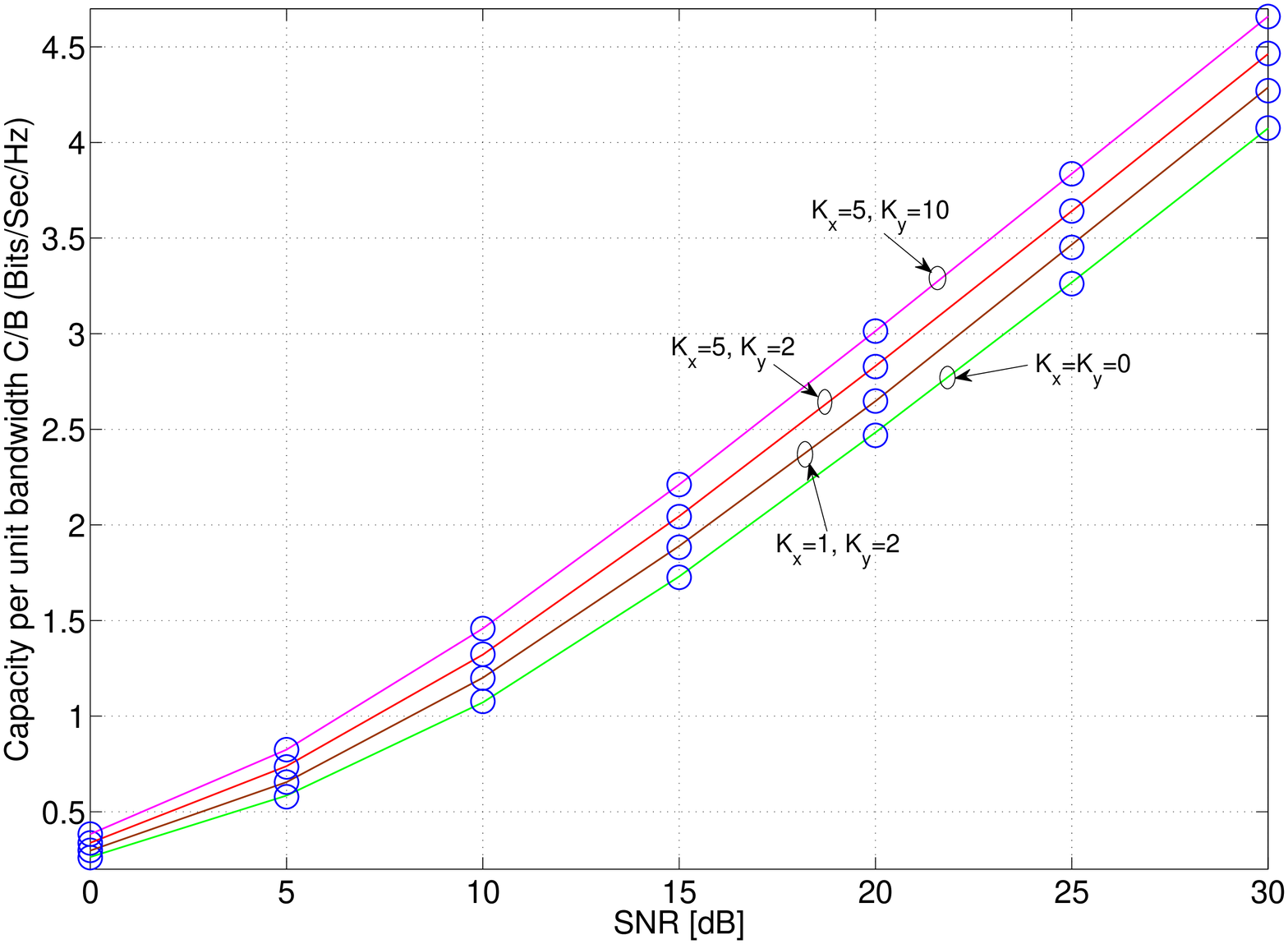,height=2.55in,width=4.0in}
    \vspace*{-0.0em}
    \mycaption{Analytical $-\!\!\!-$ and simulated $\circ$ capacity of the DF system with adaptive rate and constant power for different values of $K_x$, $K_y$, and $\bar{\gamma}_x=\bar{\gamma}_y=\bar{\gamma}$.}
    \label{fig:cap}
    \vspace*{-1.0em}
  \end{center}
\end{figure}
\begin{align}
\label{eq:firstterm}
\frac{1}{\gamma_0}\int^\infty_{\gamma_0}&f_{\gamma_z}(\gamma)d\gamma =\frac{A_xA_y}{\gamma_0}\sum^\infty_{k=0}\sum^k_{n=0}\tilde{B}_x(n)\tilde{B}_y(k-n)\nonumber\\
&\times\Gamma\left(n+1,a_x\gamma_0\right)\Gamma\left(k-n+1,a_y\gamma_0\right).
\end{align}
Further, it can be shown by using \eqref{eq:pdf_min1} and after some algebra that
\begin{align}
\label{eq:secterm}
&\int^\infty_{\gamma_0}\frac{1}{\gamma}f_{\gamma_z}(\gamma)d\gamma=A_xA_y\sum^\infty_{k=0}\sum^k_{n=0}\tilde{B}_x(n)\tilde{B}_y(k-n)\nonumber\\
&\times \Big(a^{n+1}_x(k-n)!\sum^{k-n}_{m_1=0}\!\!\frac{a^{m_1}_{y}}{m_1!}\mathcal{G}\left({n+m_1},{(a_x+a_y)\gamma_0}\right)\nonumber\\
&+a^{k-n+1}_y n! \sum^n_{m_2=0}\!\! \frac{a^{m_2}_x}{m_2!}\mathcal{G}\left({k-n+m_2},{(a_x+a_y)\gamma_0}\right)\Big),
\end{align}
where
\begin{align}
\label{eq:G}
\mathcal{G}(u,v\gamma_0)=\left\{\begin{array}{c} \frac{\Gamma\left(u,v\gamma_0\right)}{v^u},\ms \text{if}\ms u>0,\\
E_1\left(v\gamma_0\right), \ms \text{if} \ms u=0,\end{array}\right.
\end{align}
where $\Gamma\left(q,w\right)=\int^\infty_w t^{q-1}e^{-t} dt$ denotes the incomplete Gamma function and $E_1\left(w\right)=\int^\infty_1t^{-1}e^{-wt}dt, w\geq0$ is
the exponential integral. Substituting \eqref{eq:firstterm} and \eqref{eq:secterm} into \eqref{eq:optcutoff},
the optimal cutoff SNR $\gamma_0$ can be obtained numerically by using MATLAB or MATHEMATICA. Since $\int^\infty_{\gamma_0}f_{\gamma_z}(\gamma)d\gamma\leq 1$ and $\int^\infty_{\gamma_0}\left(f_{\gamma_z}(\gamma)/\gamma\right) d\gamma\geq 0$, it can be seen from \eqref{eq:optcutoff1} that $0\leq\gamma_0\leq 1$. 
\subsection{Channel Inversion with Fixed Rate}
A low complexity adaptive transmission technique is truncated channel inversion
with fixed rate where 
the transmitter only adjusts its power to maintain a
constant SNR at the destination. Truncated channel inversion
is applied if the SNR is above a cutoff $\beta_0$.
The
channel capacity in this case is given as~\cite{aloui99}\vspace*{-0.4em}
\begin{align}
\label{eq:truncatedchan}
C=0.5 B\text{log}_2\left(1+\frac{1}{\int^\infty_{\beta_0}\frac{f_{\gamma_z}(\gamma)}{\gamma}d\gamma}\right)\left(1-P_{\text{out}}\right),
\end{align}\vspace*{-0.4em}

\hspace*{-1.0em}where $P_{\text{out}}$ denotes the probability of outage given by
\begin{align}
\label{eq:prout}
P_{\text{out}}&=\text{Pr}\left(\gamma<\beta_0\right)=\int^{\beta_0}_0f_{\gamma_z}(\gamma)d\gamma=F_{\gamma_z}(\beta_0). 
\end{align}
The capacity under the channel inversion based adaptive transmission technique can be obtained by using \eqref{eq:cdf_min_final}, 
\eqref{eq:secterm}, and \eqref{eq:truncatedchan}.   
 \begin{figure}[t!]\vspace*{-1.0em}
  \begin{center}\hspace*{-2.0em}
    \psfig{figure=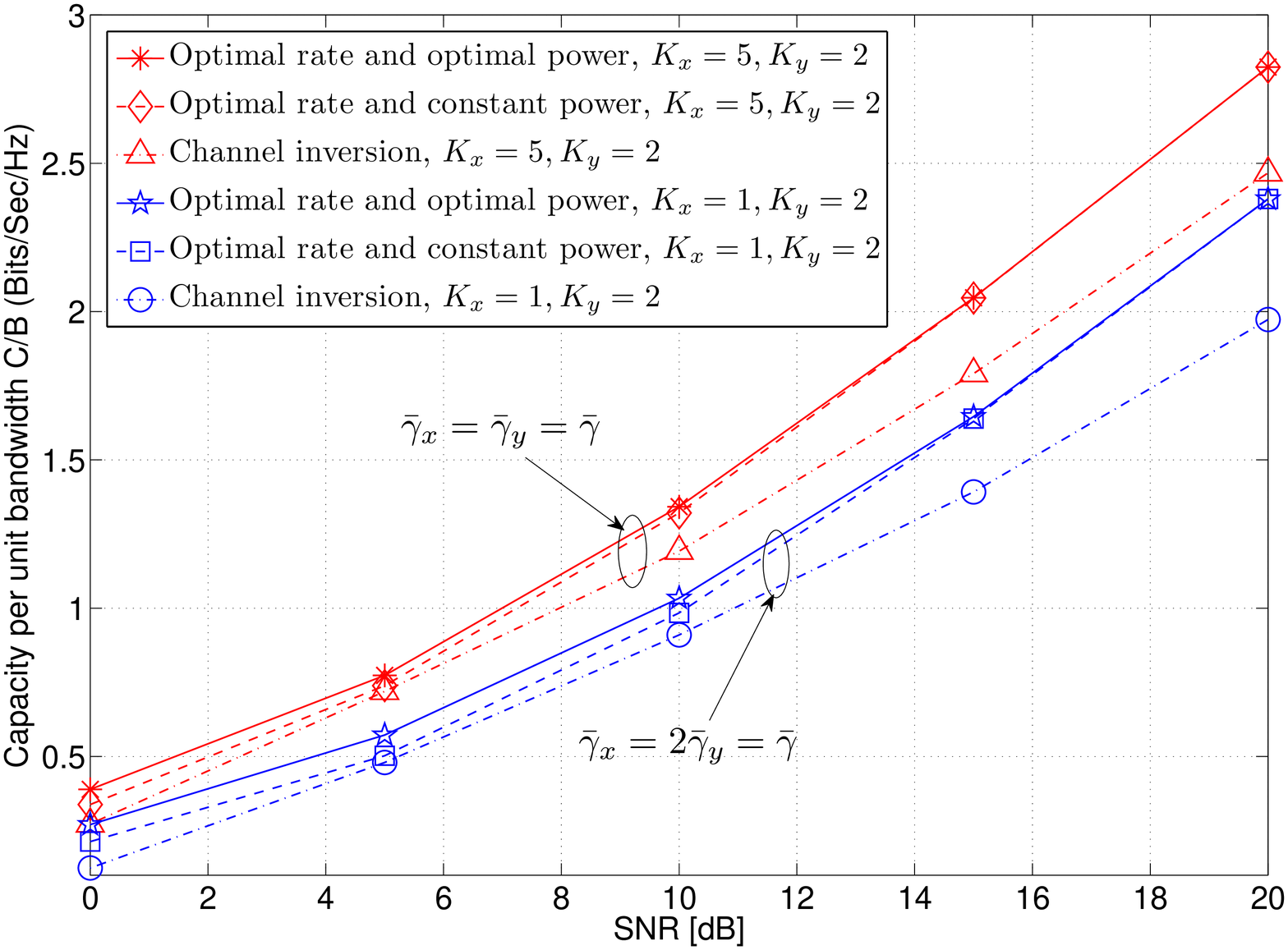,height=2.55in,width=4.0in}
    \vspace*{-0.0em}
    \mycaption{Comparison of the capacity of the DF system under different adaptive modulation schemes and with different values of $K_x$ and $K_y$.} 
    \label{fig:adp}
    \vspace*{-1.0em}
  \end{center}
\end{figure} 
\section{Numerical Results and Discussion}  
In Fig.~\ref{fig:cap}, we have plotted simulated and analytical values of the capacity of the DF system under optimal power adaptation and constant transmit power for different values of $K_x$ and $K_y$ and $\bar{\gamma}_x=\bar{\gamma}_y=\bar{\gamma}$. 
It can be seen from Fig.~\ref{fig:cap} that the simulation results match the analytical values, obtained from \eqref{eq:optimalratecap1}, closely. 
The analytical capacity of the DF system with different values of $K_x$ and $K_y$;  $\bar{\gamma}_x=\bar{\gamma}_y=\bar{\gamma}$; and $\bar{\gamma}_x=2\bar{\gamma}_y=\bar{\gamma}$ is plotted for all three adaptive modulation schemes (discussed in Section~III) in Fig.~\ref{fig:adp}. In the optimal simultaneous rate and power adaptation scheme high power
levels are assigned for good channel conditions~\cite{aloun99}; therefore, the transmit power levels are very likely to be (almost) constant
in large SNR regimes. Hence, optimal simultaneous rate and power adaptation provides (almost) no improvement in the capacity over the optimal rate adaptation and constant transmit power at high SNR values; however, a capacity gain is seen from low to moderate SNR values 
in Fig.~\ref{fig:adp}. Since the optimal rate adaptation with constant power only
adapts its rate, the DF system under Rician fading can opt for this less complexity technique than the optimal rate
and power adaptive technique at high SNR values. 
Further, the optimized capacity (with the following constraint: $\beta_0\leq 1$) of the DF Rician system with channel inversion based low complexity adaptive scheme is poorer than the other two schemes in general, as seen in Fig.~\ref{fig:adp}. 
Since $0\leq \gamma_0\leq 1$ and the probability of outage increases with increasing $\beta_0$ (as can be seen from \eqref{eq:prout}), the constraint $\beta_0\leq 1$ is used to limit the maximum outage probability of the DF system.  

The average SNR is denoted by $\bar{\gamma}$ and shown at the x-axis of Figs.~2 and 3.\vspace*{-1em} 
\appendices
\section{Proof of Lemma~\ref{lem:new1} and Lemma~\ref{lem:new}}
\label{app:new}
The CDF of $\gamma_z$ can be obtained after some algebra as follows:
\begin{align}
\label{eq:cdf_min}
F_{\gamma_z}(\gamma)
=1-\int^{\infty}_{\gamma}f_{\gamma_x}(x)dx\int^{\infty}_{\gamma}f_{\gamma_y}(y)dy.
\end{align}
From \eqref{eq:fx} and by using the series representation of the modified Bessel function~\cite[Eq. (2.3.32)]{proak08}, we can write the PDFs of $\gamma_x$ and $\gamma_y$ as 
\begin{align}
\label{eq:fxser}
f_{\gamma_x}(x)=A_xe^{-a_xx}\sum^{\infty}_{k=0}B_x(k)x^k,\nonumber\\
f_{\gamma_y}(y)=A_ye^{-a_yy}\sum^{\infty}_{k=0}B_y(k)y^k.
\end{align}\vspace*{-0.8em}

\hspace*{-0.9em}From \eqref{eq:fxser} and after some algebra, it can be shown that
\begin{align}
\label{eq:resultofintegral}
\int^{\infty}_{\gamma}f_{\gamma_x}(x)dx&=A_x\sum^\infty_{k=0}\tilde{B}_x(k)\Gamma\left(k+1,a_x\gamma\right),\nonumber\\
\int^{\infty}_{\gamma}f_{\gamma_y}(y)dy&=A_y\sum^\infty_{k=0}\tilde{B}_y(k)\Gamma\left(k+1,a_y\gamma\right).
\end{align}\vspace*{-0.8em}

\hspace*{-1.0em}By using \eqref{eq:resultofintegral} and~\cite[Eq.~(0.316)]{grand07} in \eqref{eq:cdf_min}, we get \eqref{eq:cdf_min_final}. 
After taking the derivative of $F_{\gamma_z}(\gamma)$ in \eqref{eq:cdf_min_final} with respect to $\gamma$, we have 
\begin{align}
\label{eq:pdf_min}
&f_{\gamma_z}(\gamma)=A_xA_y\sum^\infty_{k=0}\sum^k_{n=0}\tilde{B}_x(n)\tilde{B}_y(k-n)\Big(a^{n+1}_x\gamma^ne^{-a_x\gamma}\nonumber\\
&\times\Gamma\left(k-n+1,a_y\gamma\right)+a^{k-n+1}_y\gamma^{k-n}e^{-a_y\gamma}\Gamma\left(n+1,a_x\gamma\right)\Big).
\end{align}
We can further simplify \eqref{eq:pdf_min} by using the following relation: $\Gamma\left(r+1,b z\right)=e^{-bz}r!\sum^r_{m=0}\frac{b^mz^m}{m!}$,  where $r$ is a positive integer, as given in \eqref{eq:pdf_min1}.

\section{Proof of Lemma~\ref{lem:1}}
\label{app:1}
Let us denote the left hand side of \eqref{eq:lem1res} as
\begin{align}
\label{eq:1}
&\mathcal{I}=\int^\infty_0 w^\alpha e^{-\beta w} \text{ln}(1+w) dw.
\end{align} 
By substituting $1+w=t$ in the right hand side (RHS) of \eqref{eq:1}, we get
\begin{align}
\label{eq:2}
\mathcal{I}=\int^\infty_1 (t-1)^\alpha e^{-\beta (t-1)} \text{ln}t\:\: dt.
\end{align}
The following relation is obtained by using~\cite[Eq.~(1.111)]{grand07} in \eqref{eq:2}:
 \begin{align}
\label{eq:3}
\mathcal{I}=\sum^\alpha_{q=0}{}^\alpha C_q \left(-1\right)^{\alpha-q}e^\beta \int^\infty_1 t^{f(q)-1} e^{-\beta t} \text{ln}t \ms dt,
\end{align}
where $f(q)=q+1$. 
From~\cite[Eq.~(4.358.1)]{grand07} and \eqref{eq:3}, we have
 \begin{align}
\label{eq:4}
\mathcal{I}=\sum^\alpha_{q=0}{}^\alpha C_q \left(-1\right)^{\alpha-q}e^\beta \frac{\delta}{\delta f(q)}\left(\beta^{-f(q)}\Gamma\left(f(q),\beta\right)\right).
\end{align}
By using the following relation:
\begin{align}
\label{eq:5}
\frac{\delta}{\delta \nu}\Gamma\left(\nu,\mu\right)=\text{ln}\mu\Gamma\left(\nu,\mu\right)+\mu G^{3,0}_{2,3}\left(\mu\Big|\begin{array}{c}\hspace*{-1em}0,0\\-1,-1,\nu-1 \end{array}\right)
\end{align}
in \eqref{eq:4} and after some algebra we get the RHS of \eqref{eq:lem1res}.
\section{Proof of Lemma~\ref{lem:2}}
\label{app:2} 
Lemma~\ref{lem:2} can be proved by substituting $t=w/\eta$ in the left hand side of \eqref{eq:lem2res} and then using~\cite[Eq.~(4.358.1)]{grand07} and \eqref{eq:5}.  
\bibliography{IEEEabrv,biblitt}
\bibliographystyle{IEEEtran}
\end{document}